\documentclass[multphys,vecphys]{svmult}

\usepackage{makeidx}
\usepackage{graphicx}
\usepackage{multicol}
\usepackage{cite}
\usepackage[bottom]{footmisc}
\usepackage{amssymb}

\makeindex

\newcommand{\ii}{{\rm i}}

\begin{document}

\title{Resonance- and Chaos-Assisted Tunneling}

\author{Peter Schlagheck$^{1}$ \and Christopher Eltschka$^{1}$ 
  \and Denis Ullmo$^{2,3}$}

\institute{
  Institut f\"ur Theoretische Physik, Universit\"at Regensburg, 93040 Regensburg,
  Germany 
  \and
  Department of Physics, Duke University, Durham NC, USA
  \and
   Laboratoire de Physique Th\'eorique et  Mod\`eles Statistiques
   (LPTMS), 91405 Orsay Cedex, France
}

\maketitle

\begin{abstract}

We consider dynamical tunneling between two symmetry-related regular islands
that are separated in phase space by a chaotic sea.
Such tunneling processes are dominantly governed by nonlinear resonances,
which induce a coupling mechanism between ``regular'' quantum states within
and ``chaotic'' states outside the islands.
By means of a random matrix ansatz for the chaotic part of the Hamiltonian,
one can show that the corresponding coupling matrix element directly
determines the level splitting between the symmetric and the antisymmetric
eigenstates of the pair of islands.
We show in detail how this matrix element can be expressed in terms of
elementary classical quantities that are associated with the resonance.
The validity of this theory is demonstrated with the kicked Harper model. 

\end{abstract}

\section{Introduction}

Since the early days of quantum mechanics, tunneling has been recognized as
one of the hallmarks of the wave character of microscopic physics.
The possibility of a quantum particle to penetrate an energetic barrier
represents certainly one of the most spectacular implications of quantum
theory and has lead to various applications in atomic and molecular physics as
well as in mesoscopic science.
Typical scenarios in which tunneling manifests are the escape of a quantum
particle from a quasi-bound region, the transition between two or more
symmetry-related, but classically disconnected wells (which we shall focus on
in the following), as well as scattering or transport through potential
barriers.
The spectrum of scenarios becomes even richer when the concept of tunneling is
generalized to any kind of classically forbidden transitions in phase space,
i.e.\ to transitions that are not necessarily inhibited by static potential
barriers but by some other constraints of the underlying classical dynamics
(such as integrals of motion).
Such ``dynamical tunneling'' processes arise frequently in molecular systems
\cite{DavHel81JCP} and were recently realized with cold atoms propagating
in periodically modulated optical lattices \cite{HenO01N,SteOskRai01S}.

Despite its genuinely quantal nature, tunneling is strongly influenced, if not
entirely governed, by the structure of the underlying classical phase space
(see Ref.~\cite{Cre98} for a review).
This is best illustrated within the textbook example of a one-dimensional
symmetric double-well potential.
In this simple case, the eigenvalue problem can be straightforwardly solved
with the standard Wentzel-Kramers-Brillouin (WKB) ansatz \cite{LanLifQ}.
The eigenstates of this system are, below the barrier height, obtained by the
symmetric and antisymmetric linear combination of the local ``quasi-modes''
(i.e., of the wave functions that are semiclassically constructed on the
quantized orbits within each well, without taking into account the classically
forbidden coupling between the wells), and the splitting of their energies is
given by an expression of the form
\begin{equation}
  \Delta E = \frac{\hbar \omega}{\pi} \exp\left[ - \frac{1}{\hbar} \int \sqrt{2m(V(x) - E)} dx
  \right] \, . \label{eq:1dsplit}
\end{equation}
Here $E$ is the mean energy of the doublet, $V(x)$ represents the double well
potential, $m$ is the mass of the particle, $\omega$ denotes the oscillation
frequency within each well, and the integral in the exponent is performed over
the whole classically forbidden domain, i.e.\ between the inner turning points
of the orbits in the two wells.
Preparing the initial state as one of the quasi-modes (i.e., as the
even or odd superposition of the symmetric and the antisymmetric eigenstate),
the system will undergo Rabi oscillations between the wells with the frequency
$\Delta E / \hbar$.
The ``tunneling rate'' of this system is therefore given by the splitting
(\ref{eq:1dsplit}) and decreases, keeping all classical parameters fixed,
exponentially with $1 / \hbar$, what gives rise to the statement that tunneling
``vanishes'' in the classical limit.

The above expression for the splitting can also be derived in a
geometric way which is independent of the particular representation of
the phase space.  
For this purpose, it is necessary to realize that the two symmetric wells are
connected in the {\em complexified} classical phase space.
This is most conveniently expressed in terms of the local action-angle
variables $(I,\theta)$ of the well:
describing the quantized torus with the action variable $I=I_n$ in the left
well by $(p_L(I_n,\theta),q_L(I_n,\theta))$ and its counterpart in the right well by
$(p_R(I_n,\theta),q_R(I_n,\theta))$ ($p$ and $q$ are the position and momentum
variables), one can show that the analytic continuations of the two manifolds
$(p_{L/R}(I_n,\theta),q_{L/R}(I_n,\theta))$ coincide when $\theta$ is permitted to assume
complex values \cite{rem_compl}. 
Generalizing standard semiclassical theory \cite{MasFed} to this complex
Lagrangian manifold allows one to reproduce Eq.~(\ref{eq:1dsplit}), where the
exponent now contains the imaginary part of the action integral $\int p dq$ along
a path that connects the two tori in complex phase space \cite{Cre94JPA}.

This approach can be generalized to multidimensional, even non separable
systems, as long as their classical dynamics is still integrable
\cite{Cre94JPA}.
It breaks down, however, as soon as a non integrable perturbation is added to
the system (e.g.\ if the one-dimensional double-well potential is exposed to a
periodically time-dependent driving).
In that case, invariant tori may, for weak perturbations, still exist due to
the Kolmogorov-Arnol'd-Moser (KAM) theorem.
It can be shown, however, that their analytic continuation to complex
(multidimensional) angle domain generally encounters a natural boundary
in form of weak singularities that arise at a given value of the imaginary
part of the angle variable \cite{GrePer81PhD}.
Only in very exceptional cases, the complex manifolds that originate from the
two symmetry-related tori happen to meet in form of an intersection
\cite{CreFin01JPA}.
In such a situation, one can apply a semiclassical method introduced by
Wilkinson, which leads to an expression of the type (\ref{eq:1dsplit}) for the
splitting with a different $\hbar$-dependence in the prefactor \cite{Wil86PhD}.

Indeed, numerical calculations of model systems in the early nineties clearly
indicated that tunneling in non integrable systems is qualitatively different
from the above one-dimensional case.
If the perturbation introduces an appreciable chaotic layer around the
separatrix between the two wells (which in the Poincar\'e surface of section
might still be located far away from the quantized tori), the tunnel
splittings generally become strongly enhanced compared to the integrable
limit.
Moreover, they do no longer follow a smooth exponential scaling with $1 / \hbar$
as expressed by Eq.~(\ref{eq:1dsplit}), but display huge, quasi-erratic
fluctuations at variations of $\hbar$ or any other parameter of the system
\cite{LinBal90PRL,BohO93NP}.

These phenomena are traced back to the specific role that {\em chaotic}
states play in such systems
\cite{BohTomUll93PR,TomUll94PRE,DorFri95PRL,FriDor98PRE}.
In contrast to the integrable case, the tunnel doublets of the localized
quasi-modes are, in a mixed regular-chaotic system, no longer isolated in the
spectrum, but resonantly interact with states that are associated with the
chaotic part of phase space.
Due to their delocalized nature, such chaotic states typically exhibit a
significant overlap with the boundary regions of both regular wells.  
They may therefore provide an efficient coupling mechanism between the
quasi-modes -- which becomes particularly effective whenever one of the
chaotic levels is shifted exactly on resonance with the tunnel doublet.
This coupling mechanism generally enhances the tunneling rate, but may 
accidentally also lead to a complete {\em suppression} thereof, arising at
very specific values of $\hbar$ or other parameters
\cite{GroO91PRL,AveOsoMoi02PRL}.

The validity of this ``chaos-assisted'' tunneling picture was essentially
confirmed by a simple statistical ansatz in which the quantum dynamics within
the chaotic part of the phase space was represented by a random matrix from
the Gaussian orthogonal ensemble (GOE)
\cite{BohTomUll93PR,TomUll94PRE,LeyUll96JPA}.
In presence of small coupling coefficients between the regular states and the
chaotic domain, this random matrix ansatz yields a truncated Cauchy
distribution for the probability density to obtain a level splitting of the
size $\Delta E$.
Such a distribution is indeed encountered in the exact quantum splittings,
which was demonstrated for the two-dimensional quartic oscillator
\cite{LeyUll96JPA} as well as, later on, for the driven pendulum Hamiltonian
that describes the tunneling process of cold atoms in periodically modulated
optical lattices \cite{MouO01PRE,MouDel03PRE}.
The random matrix ansatz can be straightforwardly generalized to the
tunneling-induced decay of quasi-bound states in open systems, which is
relevant for the ionization process of non dispersive electronic wave packets
in resonantly driven hydrogen atoms \cite{ZakDelBuc98PRE}.
Chaos-assisted tunneling is, furthermore, not restricted to quantum mechanics,
but arises also in the electromagnetic context, as was shown by experiments on
optical cavities \cite{NoeSto97N} and microwave billiards
\cite{DemO00PRL,HofO05PRE}.

A quantitative prediction of the {\em average} tunneling rate was not
possible in the above-mentioned theoretical works.  As we shall see
later on, this average tunneling rate is directly connected to the
coupling matrix element between the regular and the chaotic states,
and the strength of this matrix element was unknown and introduced in
an ad-hoc way.
A natural way to tackle this problem would be to base the semiclassical
description of the chaos-assisted tunneling process on {\em complex}
trajectories \cite{MilGeo72JCP}, to be obtained, e.g., by solving Hamilton's
equations of motion along complex time paths.
Such trajectories would possibly involve a classically forbidden escape out of
(and re-entrance into) the regular islands as well as classically allowed
propagation within the chaotic sea.

For the case of time-dependent propagation processes, such as the
evolution of a wave packet that was initially confined to a regular
region, this ambitious semiclassical program can be carried out in a
comparatively straightforward way, which is nevertheless hard to
implement in practice \cite{ShuIke95PRL,ShuIke96PRL}.  Indeed, Shudo,
Ikeda and coworkers showed in this context that the selection of
complex paths that contribute to the semiclassical propagator requires
a careful consideration of the Stokes phenomenon, in order to avoid
``forbidden'' trajectories that would lead to an exponential increase
(instead of decrease) of the tunneling amplitude
\cite{ShuIke95PRL,ShuIke96PRL}.  This semiclassical method can be
generalized to scattering problems in presence of non integrable
barriers \cite{OniO01PRE,TakIke03JPA,TakIke05EPL} and allows one to
interpret structures in the tunneling tail of the wave function (such
as plateaus and ``cliffs'') in terms of chaos in the complex classical
domain.

A crucial step towards the semiclassical treatment of
``time-independent'' tunneling problems, such as the determination of
the level splitting between nearly degenerate states in classically
disconnected wells, was undertaken by Creagh and Whelan.  They showed
that the splitting-weighted density of states $\sum_n \Delta E_n
\delta(E - E_n)$ (where $E_n$ are the mean energies and $\Delta E_n$
the splittings of the doublets) can be expressed as a Gutzwiller-like
trace formula involving complex periodic orbits that connect the two
wells \cite{CreWhe96PRL,CreWhe99PRL}.  Such orbits also permit to
determine the individual splittings $\Delta E_n$, provided the
wave functions of the associated quasimodes are known
\cite{CreWhe99AP}.  In practice, this approach can be successfully
applied to fully chaotic wells that are separated by an energetic
barrier \cite{CreWhe96PRL,CreWhe99PRL,CreWhe00PRL}, since in such
systems the semiclassical tunneling process is typically dominated by
a single instanton-type orbit.  A generalization to chaos-assisted
tunneling, where many different orbits would be expected to
contribute, does not seem straightforward.

In time-dependent propagation problems as well as the in approaches based on
the Gutzwiller trace formula \cite{Gut}, the Lagrangian manifolds that need to
be constructed generally have simple analytical structures.
This is the case for time-dependent problems because the initial wave function
is usually chosen as a Gaussian wave packet or a plane wave, which are
semiclassically associated with simple manifolds.
In Gutzwiller-like approaches, this due to the inherent structure of the
theory, but at the cost of involving long orbits when individual eigenlevels
need to be described.
For the problem of determining individual level splittings in chaos-assisted
tunneling, however, the natural objects with which one would have to start
with are the {\em invariant tori} of the system, the analytical structure of
which is extremely involved in case of a mixed regular-chaotic dynamics.
As was mentioned above, these tori cannot be analytically continued very far
in the complex phase space as a natural boundary is usually encountered.
As a consequence, it does not seem that a description in terms of complex
dynamics could be easily developed for chaos-assisted tunneling.

The main purpose of our contribution is to show that a relatively
comprehensive description of the tunneling between two regular island in a
mixed phase space can nevertheless be obtained.
This description involves a number of approximations which might be improved to
reach better accuracy.
It however leads to quantitative predictions for the tunneling rates which are
in sufficiently good agreement with the exact quantum data to ensure that the 
{\em mechanism} that underlies this process is correctly accounted for.
We shall, in this context, particularly focus on the classically forbidden
transition from the regular island into the chaotic sea.  
As already pointed out above, the associated coupling matrix element
determines the average tunnel splittings between the islands, which
means that a simple semiclassical access to this matrix element would
open the possibility to quantitatively estimate tunneling rates in
systems with mixed dynamics.

An important first step in this direction was undertaken by Podolskiy and
Narimanov:
By assuming a perfectly clean, harmonic-oscillator like dynamics within the
regular island and a structureless chaotic sea outside the outermost invariant
torus of the island, a semiclassical expression of the form
\begin{equation}
  \Delta E \simeq \gamma \hbar \frac{\Gamma(\nu,2\nu)}{\Gamma(\nu+1,0)} \stackrel{\nu \gg 1}{\simeq} 
  \frac{\gamma \hbar}{\sqrt{2\pi \nu^3}} {\rm e}^{-(1-\ln 2) \nu} \quad 
  \mbox{with} \quad \nu = A / (\pi \hbar) \label{eq:pod}
\end{equation}
was derived for the average eigenphase splitting \cite{PodNar03PRL}.
Here $A$ is the phase space area covered by the regular island, and
$\Gamma(a,x)$ denotes the incomplete Gamma function \cite{Abr}.
The prefactor $\gamma$ is system specific, but does not depend on $\hbar$, which
permits a prediction of the general decay behavior of the splittings with 
$1 / \hbar$.
Good agreement was indeed found in a comparison with the exact splittings
between near-degenerate optical modes that are associated with a pair of
symmetric regular islands in a non integrable micro-cavity \cite{PodNar03PRL}
(see also Ref.~\cite{PodNar05OL}).

The theory was furthermore applied to the dynamical tunneling process in
periodically modulated optical lattices, for which the splittings between the
left- and the right-moving stable eigenmodes were calculated in
Ref.~\cite{MouO01PRE}.
Those splittings seem to be very well described by Eq.~(\ref{eq:pod}) for low
and moderate values of $1 / \hbar$, but display significant deviations from this
expression deeper in the semiclassical regime \cite{PodNar03PRL}.
Indeed, the critical value of $1 / \hbar$ beyond which this disagreement occurs
coincides with the edge of an intermediate {\em plateau} in the splittings,
which extends over the rather large, ``macroscopic'' range $10 <  1 / \hbar < 30$.
As we shall see in Section \ref{s:kh}, such prominent plateau structures
appear quite commonly in chaos-assisted tunneling processes (see also
Ref.~\cite{RonO94PRL}), and a smooth, monotonously decreasing expression of
the type (\ref{eq:pod}) cannot account for them.

To understand the origin of such plateaus, it is instructive to step back to
the conceptually simpler case of {\em nearly integrable} dynamics, where the
perturbation from the integrable Hamiltonian is sufficiently small such that
macroscopically large chaotic layers are not yet developed in the Poincar\'e
surface of section.
In such systems, the main effect of the perturbation consists in the
manifestation of chain-like substructures in the phase space, which arise at
{\em nonlinear resonances} between the eigenmodes of the unperturbed
Hamiltonian, or, in periodically driven systems, between the external driving
and the unperturbed oscillation within the well.
In a similar way as for the quantum pendulum Hamiltonian, such resonances
induce additional tunneling paths in the phase space, which lead to
couplings between states that are located in the {\em same} well
\cite{Ozo84JPC,UzeNoiMar83JCP}.

The relevance of this effect for the near-integrable tunneling process between
two symmetry-related wells was first pointed out by Bonci et
al.~\cite{BonO98PRE} who argued that such resonances may lead to a strong
enhancement of the tunneling rate, due to couplings between lowly and highly
excited states within the well which are permitted by near-degeneracies in
the spectrum.
In Refs.~\cite{BroSchUll01PRL,BroSchUll02AP}, a quantitative semiclassical
theory of near-integrable tunneling was formulated on the basis of this
principal mechanism.
This theory allows one to reproduce the exact quantum splittings on the basis
of purely classical quantities that can be extracted from the phase space, and
takes into account high-order effects such as the coupling via a sequence of
different resonance chains \cite{BroSchUll01PRL,BroSchUll02AP}.
Recent studies by Keshavamurthy on classically forbidden coupling processes in
model Hamiltonians that mimic the dynamics of simple molecules confirm that
the ``resonance-assisted'' tunneling scenario prevails not only in
one-dimensional systems that are subject to a periodic driving (such as the
``kicked Harper'' model which was studied in
Ref.~\cite{BroSchUll01PRL,BroSchUll02AP}), but also in autonomous systems with
two and even three degrees of freedom \cite{Kes05JCP,Kes05XXX}.

Our main focus in this review is that such nonlinear resonances play an
equally important role also in the mixed-regular chaotic case.
Indeed, they have recently been shown to be primarily responsible for the
coupling between the regular island and the chaotic sea in the semiclassical
regime \cite{EltSch05PRL}.  
In combination with the above random matrix ansatz for the chaotic states, a
simple analytical expression for the average tunneling rate is obtained in
this way, which provides a straightforward interpretation of the plateau
structures in the splittings in terms of multi-step coupling processes induced
by the resonances.

To explain this issue in more detail, we start, in Section \ref{s:th}, with a
description of the effect of nonlinear resonances on the quantum dynamics
within a regular region in phase space.
We then discuss how the presence of such resonances leads to a modification of
the tunnel coupling in the near-integrable as well as in the mixed
regular-chaotic regime, and present a simple semiclassical scheme that allows
one to reproduce the associated tunneling rates. 
Applications to tunneling processes in the kicked Harper model are studied in
Section \ref{s:kh}.

\section{Theory of resonance-assisted tunneling}

 \label{s:th}

We restrict our study to systems with one degree of freedom that evolve under
a periodically time-dependent Hamiltonian $H(p,q,t) = H(p,q,t + \tau)$.
We suppose that, for a suitable choice of internal parameters, the classical
phase space of $H$ is mixed regular-chaotic and exhibits two symmetry-related
regular islands that are embedded into the chaotic sea.
This phase space structure is most conveniently visualized by a stroboscopic
Poincar\'e section, where $p$ and $q$ are plotted at the times 
$t = n \tau (n \in \mathbb{Z})$.
Such a Poincar\'e section typically reveals the presence of chain-like
substructures within the regular islands, which arise due to 
nonlinear resonances between the external driving and the internal
oscillation around the island's center.
We shall assume now that the two islands exhibit a prominent $r$:$s$ resonance
--- i.e., where $s$ internal oscillation periods match $r$ driving periods,
and $r$ sub-islands are visible in the stroboscopic section.

The classical motion in the vicinity of the $r$:$s$ resonance is approximately
integrated by secular perturbation theory \cite{LicLie} (see also
Ref.~\cite{BroSchUll02AP}).
For this purpose, we formally introduce a time-independent Hamiltonian
$H_0(p,q)$ that approximately reproduces the regular motion in the islands
and preserves the discrete symmetry of $H$.
The phase space generated by this integrable Hamiltonian consequently exhibits
two symmetric wells that are separated by an energetic barrier and ``embed''
the two islands of $H$.
In terms of the action-angle variables $(I,\theta)$ describing the dynamics within
each of the wells, the total Hamiltonian can be written as 
\begin{equation}
  H(I,\theta,t) = H_0(I) + V(I,\theta,t)
\end{equation}
where $V$ would represent a weak perturbation in the center of the island.

The nonlinear $r$:$s$ resonance occurs at the action variable $I_{r:s}$ that
satisfies the condition
\begin{equation}
  r \Omega_{r:s} = s \frac{2\pi}{\tau}
 \quad \mbox{with} \quad 
  \Omega_{r:s} \equiv \left. \frac{d H_0}{d I}\right|_{I=I_{r:s}} \, .
\end{equation}
We now perform a canonical transformation to the frame that corotates with
this resonance.
This is done by leaving $I$ invariant and modifying $\theta$ according to
\begin{equation}
  \theta \mapsto \vartheta = \theta - \Omega_{r:s} t \, .
\end{equation}
This time-dependent shift is accompanied by the transformation
$H \mapsto \mathcal{H} = H - \Omega_{r:s} I$ in order to ensure that the new corotating
angle variable $\vartheta$ is conjugate to $I$.
The motion of $I$ and $\vartheta$ is therefore described by the new Hamiltonian
\begin{equation}
  \mathcal{H}(I,\vartheta,t) = \mathcal{H}_0(I) + \mathcal{V}(I,\vartheta,t)
\end{equation}
with
\begin{eqnarray}
  \mathcal{H}_0(I) & = & H_0(I) - \Omega_{r:s} I \, , \\
  \mathcal{V}(I,\vartheta,t) & = & V(I,\vartheta + \Omega_{r:s} t,t) \, . \label{eq:Vrot}
\end{eqnarray}

The expansion of $\mathcal{H}_0$ in powers of $I - I_{r:s}$ yields
\begin{equation}
  \mathcal{H}_0(I) \simeq \mathcal{H}_0^{(0)} + \frac{(I - I_{r:s})^2}{2 m_{r:s}} +
  \mathcal{O}\left((I - I_{r:s})^3\right) \label{eq:H0res}
\end{equation}
with a constant $\mathcal{H}_0^{(0)}$ and a quadratic term that is
characterized by the effective ``mass'' parameter $m_{r:s}$.
Hence, $d \mathcal{H}_0 / d I$ is comparatively small for $I \simeq I_{r:s}$, which
implies that the corotating angle $\vartheta$ varies slowly in time near the
resonance.
This justifies the application of adiabatic perturbation theory \cite{LicLie},
which effectively amounts, in first order, to replacing $\mathcal{V}(I,\vartheta,t)$
by its time average over $r$ periods of the driving (using the fact that 
$\mathcal{V}$ is periodic in $t$ with the period $r \tau$) \cite{rem_ad}.
By making a Fourier series expansion for $V(I,\theta,t)$ in both $\theta$ and $t$, one
can show that the resulting time-independent perturbation term is
($2\pi/r$)-periodic in $\vartheta$.
We therefore obtain, after this transformation, the time-independent
Hamiltonian $\mathcal{H}_0(I) + V_{\rm av}(I,\vartheta)$ 
where $V_{\rm av}$ can be written as the Fourier series
\begin{equation}
  V_{\rm av}(I,\vartheta) \equiv \frac{1}{r\tau}\int_0^{r\tau} \mathcal{V}(I,\vartheta,t) 
  = \sum_{k=0}^{\infty} V_k(I)\cos(k r \vartheta + \phi_k) \, . \label{eq:Vres}
\end{equation}
This effective Hamiltonian is further simplified by neglecting the action
dependence of the Fourier coefficients of $V_{\rm av}$ --- i.e., we use 
$V_k \equiv V_k(I=I_{r:s})$ in Eq.~(\ref{eq:Vres}) --- and by employing the
quadratic approximation (\ref{eq:H0res}) of $H_0(I)$ around $I=I_{r:s}$.
Leaving out constant terms, we finally obtain the effective integrable
Hamiltonian 
\begin{equation}
  H_{\rm eff}(I,\vartheta) = \frac{(I - I_{r:s})^2}{2 m_{r:s}} + 
  \sum_{k=1}^\infty V_k \cos(k r \vartheta + \phi_k) \, . \label{heff}
\end{equation}

The structure of this Hamiltonian exhibits all ingredients that are necessary
to understand the enhancement of tunneling.
This can be most explicitly seen by applying quantum perturbation theory to
the semiclassical quantization of $H_{\rm eff}$, treating the ``kinetic'' term
$\propto (\hat{I} - I_{r:s})^2$ as unperturbed part ($\hat{I} \equiv - \ii \hbar \partial/\partial \vartheta$) and the
``potential'' term  $\sum_{k=1}^\infty V_k \cos(k r \hat{\vartheta} + \phi_k)$ as perturbation:
Within the unperturbed eigenbasis $\langle\vartheta|n\rangle \sim \exp(\ii n \vartheta)$ (i.e., the
eigenbasis of $H_0$ in action-angle variables), couplings are introduced
between the states $|n\rangle$ and $|n+kr\rangle$ according to
\begin{equation}
  \langle n+kr|H_{\rm eff}|n\rangle = \frac{1}{2} V_k e^{\ii \phi_k} \, .
\end{equation}
As a consequence, the ``true'' eigenstates $|\psi_n\rangle$ of $H_{\rm eff}$ contain
admixtures from unperturbed modes $|n'\rangle$ that satisfy the selection rule
$|n' - n| = kr$ with integer $k$.
They are given by the perturbative expression
\begin{eqnarray}
  |\psi_n\rangle & = & |n\rangle + \sum_{k\neq0} \frac{V_k e^{\ii \phi_k} / 2}{E_n - E_{n+kr}} |n+kr\rangle + \nonumber \\
  & & + \sum_{k,k'\neq0} \frac{V_k e^{\ii \phi_k} /2}{E_n - E_{n+kr}} 
  \frac{V_{k'} e^{\ii \phi_{k'}} /2}{E_n - E_{n+kr+k'r}}|n+kr+k'r\rangle + \ldots \label{npert}
\end{eqnarray}
where $E_n$ denote the unperturbed eigenenergies of $H_{\rm eff}$, i.e., the
eigenvalues of $H_0(I) - \Omega_{r:s} I$.

Within the quadratic approximation of $H_0(I)$ around $I_{r:s}$, we obtain from
Eq.~(\ref{heff})
\begin{equation}
  E_n = \frac{(I_n - I_{r:s})^2}{2m_{r:s}} \label{eq:en}
\end{equation}
where the quantized actions are given by
\begin{equation}
  I_n = \hbar ( n + 1/2)
\end{equation}
(taking into account the generic Maslov index $\mu=2$ for regular islands).
This results in the energy differences
\begin{equation}
  E_n - E_{n'} = \frac{1}{2m_{r:s}} (I_n - I_{n'}) (I_n + I_{n'} - 2 I_{r:s})
  \, . \label{Ediff}
\end{equation}
>From this expression, we see that the admixture between $|n\rangle$
and $|n'\rangle$ becomes particularly strong if the $r$:$s$ resonance
is symmetrically located between the two tori that are associated with
the actions $I_n$ and $I_{n'}$ --- i.e., if $I_n + I_{n'} \simeq 2
I_{r:s}$.  The presence of a significant nonlinear resonance within a
region of regular motion provides therefore an efficient mechanism to
couple the local ``ground state'' --- i.e, the state that is semiclassically
localized in the center of that region (with action variable $I_0 < I_{r:s}$)
--- to a highly excited state (with action variable $I_{kr} > I_{r:s}$).

It is instructive to realize that the Fourier coefficients $V_k$ of the
perturbation operator decrease rather rapidly with increasing $k$.
Indeed, one can derive under quite general circumstances the asymptotic
scaling law
\begin{equation}
  V_k \sim (kr)^\gamma V_0 \exp[ - k r \Omega_{r:s} t_{\rm im}(I_{r:s}) ] \label{Vk}
\end{equation}
for large $k$, which is based on the presence of singularities of the
complexified tori of the integrable approximation $H_0(I)$
\cite{BroSchUll02AP}.
Here $t_{\rm im}(I)$ denotes the imaginary time that elapses from the (real) torus with
action $I$ to the nearest singularity in complex phase space, $\gamma$ corresponds
to the degree of the singularity, and $V_0$ contains information about the
corresponding residue near the singularity as well as the strength of the
perturbation.
The expression (\ref{Vk}) is of little practical relevance as far as the
concrete determination of the coefficients $V_k$ is concerned.
It permits, however, to estimate the relative importance of different
perturbative pathways connecting the states $|n\rangle$ and $|n+kr\rangle$ in
Eq.~(\ref{npert}).
Comparing e.g.\ the amplitude $\mathcal{A}_2$ associated with a single step
from $|n\rangle$ to $|n+2r\rangle$ via $V_2$ and the amplitude $\mathcal{A}_1$ associated
with two steps from $|n\rangle$ to $|n+2r\rangle$ via $V_1$, we obtain from
Eqs.~(\ref{Ediff}) and (\ref{Vk}) the ratio
\begin{equation}
  \mathcal{A}_2 / \mathcal{A}_1 \simeq  \frac{2^\gamma r^{2-\gamma} \hbar^2}{m_{r:s} V_0} 
	e^{\ii (\phi_2 - 2 \phi_1)}
\end{equation}
under the assumption that the resonance is symmetrically located in between
the corresponding two tori.
Since $V_0$ can be assumed to be finite in mixed regular-chaotic systems, we
infer that the second-order process via the stronger coefficient $V_1$ will
more dominantly contribute to the coupling between $|n\rangle$ and $|n+2r\rangle$ in the
semiclassical limit $\hbar \to 0$.

\begin{figure}[t]
  \begin{center}
    \includegraphics*[width=0.5\textwidth]{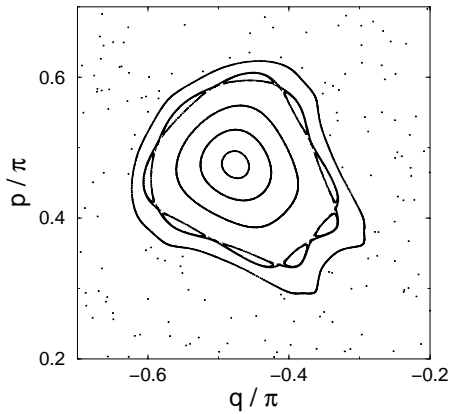}
  \end{center}
  \caption{
    Classical phase space of the kicked Harper Hamiltonian (at $\tau = 3$),
    showing a regular island with an embedded 9:8 resonance.
    The thick solid line represents the ``outer'' and the thin dashed line the
    ``inner'' separatrix of the resonance.
    \label{fg:sep}
  }
\end{figure}

A similar result is obtained from a comparison of the one-step process via
$V_k$ with the $k$-step process via $V_1$, where we again find that the latter
more dominantly contributes to the coupling between $|n\rangle$ and $|n+kr\rangle$ in the
limit $\hbar \to 0$.
We therefore conclude that in mixed regular-chaotic systems the semiclassical
tunneling process can be adequately described by an effective pendulum-like
Hamiltonian in which the Fourier components $V_k$ with $k > 1$ are completely
neglected:
\begin{equation}
  H_{\rm eff}(I,\vartheta) = \frac{(I - I_{r:s})^2}{2 m_{r:s}} + 2 V_{r:s} \cos r \vartheta
  \label{pend}
\end{equation}
with $2 V_{r:s} \equiv V_1$ \cite{EltSch05PRL} (we assume $\phi_1 = 0$ without loss of
generality).
This simple form of the effective Hamiltonian allows us to determine the
parameters $I_{r:s}$, $m_{r:s}$ and $V_{r:s}$ from the Poincar{\'e} map of the
classical dynamics, without explicitly using the transformation to the
action-angle variables of $H_0$.
To this end, we numerically calculate the monodromy matrix
$M_{r:s} \equiv \partial(p_f,q_f) / \partial(p_i,q_i)$ of a stable periodic point of the
resonance (which involves $r$ iterations of the stroboscopic map) as well as
the phase space areas $S_{r:s}^+$ and $S_{r:s}^-$ that are enclosed by the
outer and inner separatrices of the resonance, respectively (see also
Fig.~\ref{fg:sep}).
Using the fact that the trace of $M_{r:s}$ as well as the phase space areas
$S_{r:s}^\pm$ remain invariant under the canonical transformation to $(I,\vartheta)$, we
infer
\begin{eqnarray}
  I_{r:s} & = & \frac{1}{4 \pi} ( S_{r:s}^+ + S_{r:s}^- ) \, , \label{eq:area} \\
  \sqrt{2 m_{r:s} V_{r:s}} & = & \frac{1}{16} ( S_{r:s}^+ - S_{r:s}^- ) \, ,
  \label{eq:sep} \\
  \sqrt{\frac{2 V_{r:s}}{m_{r:s}}} & = & \frac{1}{r^2 \tau} \arccos({\rm tr} \,
  M_{r:s}/2) \label{eq:trm}
\end{eqnarray}
from the integration of the dynamics generated by $H_{\rm eff}$.

Quantum mechanically, the tunneling process between the $n$th excited
quantized torus and its counterpart in the symmetry-related island manifests
itself in a small level splitting between the associated symmetric and
antisymmetric eigenstates.
In our case of a periodically driven system with one degree of freedom, these
eigenstates arise from a diagonalization of the unitary time evolution
operator $U$ over one period $\tau$ of the driving, and the splitting is defined
by the difference
\begin{equation}
  \Delta \varphi_n = |\varphi_n^+ - \varphi_n^-|
\end{equation}
between the corresponding eigenphases $\varphi_n^\pm$ of the symmetric and
antisymmetric state.
In the integrable limit, these eigenphase splittings are trivially related to
the energy splittings $\Delta E_n^{(0)}$ of the unperturbed Hamiltonian $H_0$ via 
\begin{equation}
  \Delta \varphi_n^{(0)} = \tau \Delta E_n^{(0)} / \hbar \, .
\end{equation}
The latter can be semiclassically calculated by the analytic continuation of
the tori to the complex domain \cite{Cre94JPA}, and are given by
\begin{equation}
  \Delta E_n^{(0)} = \frac{\hbar \Omega_n}{\pi} \exp( - \sigma_n / \hbar ) \label{split0}
\end{equation}
where $\Omega_n$ is the oscillation frequency of the $n$th quantized torus and
$\sigma_n$ denotes the imaginary part of the action integral along the complex path
that joins the two symmetry-related tori.

In the near-integrable case \cite{BroSchUll01PRL,BroSchUll02AP}, the presence
of a prominent $r$:$s$ resonance provides an efficient coupling between the
ground state and highly excited states within the regular region.
Within perturbation theory, we obtain for the modified ground state splitting
\begin{equation}
  \Delta \varphi_0 = \Delta \varphi_0^{(0)} + \sum_k |\mathcal{A}_{kr}^{(r:s)}|^2 \Delta \varphi_{kr}^{(0)} \label{split1}
\end{equation}
where $\mathcal{A}_{kr}^{(r:s)} = \langle kr | \psi_0 \rangle$ denotes the admixture of the
$(kr)$th excited component to the perturbed ground state according to
Eq.~(\ref{npert}).
The rapid decrease of the amplitudes $\mathcal{A}_{kr}^{(r:s)}$ with $k$ is
compensated by an exponential increase of the unperturbed splittings 
$\Delta \varphi_{kr}^{(0)}$, arising from the fact that the tunnel action $\sigma_n$ in
Eq.~(\ref{split0}) generally decreases with increasing $n$.
The maximal contribution to the modified ground state splitting is generally
provided by the state $| kr \rangle$ for which $I_{kr} + I_0 \simeq 2 I_{r:s}$ --- i.e.,
which in phase space is most closely located to the torus that lies
symmetrically on the opposite side of the resonance chain.
This contribution is particularly enhanced by a small energy denominator
(see Eq.~(\ref{Ediff})) and typically dominates the sum in Eq.~(\ref{split1}).

In the mixed regular-chaotic case, invariant tori exist only up to a maximum
action variable $I_c$ corresponding to the outermost boundary of the regular
island in phase space.
Beyond this outermost invariant torus, multiple overlapping resonances provide
various couplings and pathways such that unperturbed states in this regime
can be assumed to be strongly connected to each other.
Under such circumstances, the classically forbidden coupling between the two
symmetric islands does not require any ``direct'' tunneling process of the
type (\ref{split0});
it can be achieved by a resonance-induced transition from the ground state to
a state within the chaotic domain.

\begin{figure}[t]
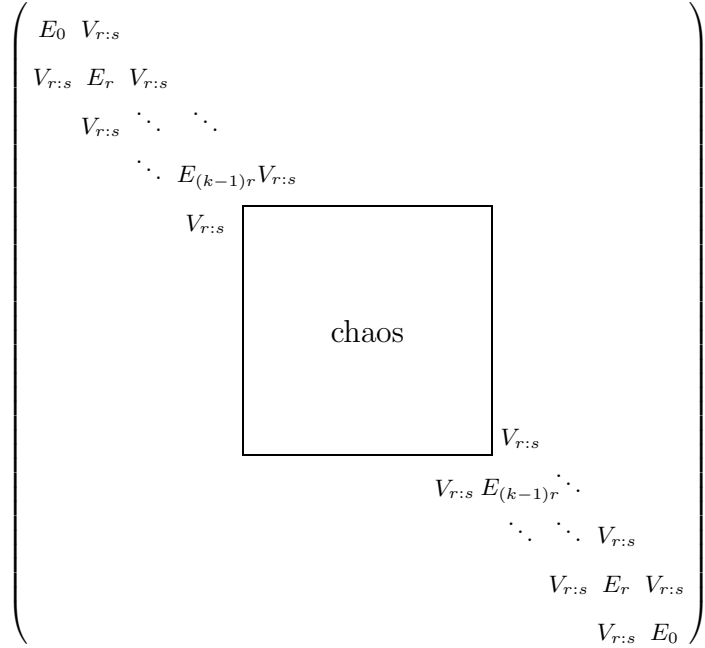

  \begin{displaymath}
    \newlength{\Vrs}
    \settowidth{\Vrs}{$V_{r:s}$}
    \newlength{\csize}
    \setlength{\csize}{9.5em}
    \left(
      \begin{array}{cccclcccc}
        E_0 & V_{r:s} & \phantom{\ddots}\\
        V_{r:s} & E_r & V_{r:s} & \phantom{\ddots}\\
        & V_{r:s} & \ddots & \ddots \\
        & & \ddots & E_{(k-1)r} \hspace*{-0.3cm} 
        & \hspace*{0.2cm} V_{r:s} \\[2mm]
        & & & \parbox[c][\csize][t]{\Vrs}{$V_{r:s}$} &
        \framebox{\parbox[c][\csize][c]{\csize}{\centering\large chaos}}
        & \parbox[l][\csize][b]{\Vrs}{$V_{r:s}$}\\
        & & & & \parbox{\csize}{\hfill $V_{r:s}$} & 
        \hspace*{-0.3cm} E_{(k-1)r} \hspace*{-0.3cm} & \ddots\\
        & & & & & \ddots & \ddots & V_{r:s}\\
        & & & & & \phantom{\ddots} & V_{r:s} & E_r & V_{r:s}\\
        & & & & & & \phantom{\ddots} & V_{r:s} & E_0
      \end{array}
    \right)
  \end{displaymath}
  \caption{Sketch of the effective Hamiltonian matrix that describes tunneling
    between the symmetric quasi-modes in the two separate regular islands.
    The regular parts (upper left and lower right band) includes only
    components that are coupled to the island's ground state by the $r$:$s$
    resonance.
    The chaotic part (central square) consists of a full sub-block with
    equally strong couplings between all basis states with actions beyond the
    outermost invariant torus of the islands.
    \label{fg:heff}
  }
\end{figure}

The structure of the effective Hamiltonian that describes this coupling
process is depicted in Fig.~\ref{fg:heff}.
We assume here that the perturbation induced by the nonlinear $r$:$s$
resonance is adequately described by the simplified pendulum Hamiltonian
(\ref{pend}).
Separating the Hilbert space into an ``even'' and ``odd'' subspace with
respect to the discrete symmetry of $H$ and eliminating intermediate states
within the regular island leads to an effective Hamiltonian matrix of the form
\begin{equation}
  H_{\rm eff}^\pm = \left( 
    \begin{array}{ccccc}
      E_0 & V_{\rm eff} & 0 & \cdots & 0 \\
      V_{\rm eff} & H_{11}^\pm & \cdots & \cdots & H_{1N}^\pm \\
      0 & \vdots & & & \vdots \\
      \vdots & \vdots & & & \vdots \\
      0 & H_{N1}^\pm & \cdots & \cdots & H_{NN}^\pm
    \end{array}
  \right) \, .
  \label{eq:model}
\end{equation}
for each symmetry class.
The effective coupling matrix element between the ground state and the chaos
block $(H_{ij}^\pm)$ is given by 
\begin{equation}
  V_{\rm eff} = V_{r:s} \prod_{l=1}^{k - 1}\frac{V_{r:s}}{E_0 - E_{lr}} \label{eq:veff}
\end{equation}
where $E_n$ are the unperturbed energies (\ref{eq:en}) of $H_{\rm eff}$.
Here $|kr\rangle$ represents the lowest unperturbed state that is connected by the
$r$:$s$ resonance to the ground state and located outside the outermost
invariant torus of the island (i.e., $I_{(k-1)r} < I_c < I_{kr}$).

In the simplest possible approximation, which follows the lines of
Refs.~\cite{TomUll94PRE,LeyUll96JPA}, we neglect the effect of partial
barriers in the chaotic part of the phase space \cite{BohTomUll93PR} and
assume that the chaos block $(H_{ij}^\pm)$ is adequately modeled by a random
hermitian matrix from the Gaussian orthogonal ensemble (GOE).
After a pre-diagonalization of $(H_{ij}^\pm)$, yielding the eigenstates $\phi_j^\pm$ and
eigenenergies $\mathcal{E}_j^\pm$, we can perturbatively express the shifts of
the symmetric and antisymmetric ground state energies by
\begin{equation}
  E_0^\pm = E_0 + V_{\rm eff}^2 \sum_{j=1}^N \frac{|\langle kr|\phi_j^\pm\rangle|^2}{E_0 -
    \mathcal{E}_j^\pm} \, .
\end{equation}
Performing the random matrix average for the eigenvectors, we obtain 
\begin{equation}
  |\langle kr|\phi_j^\pm\rangle|^2 \simeq 1/N
\end{equation}
for all $j=1\ldots N$, which simply expresses the fact that none of the basis
states is distinguished within the chaotic block $(H_{ij})$.

As was shown in Ref.~\cite{LeyUll96JPA}, the random matrix average over the
eigenvalues $\mathcal{E}_j^\pm$ gives rise to a Cauchy distribution for the 
shifts of the ground state energies, and consequently also for the splittings
\begin{equation}
  \Delta E_0 = |E_0^+ - E_0^-|
\end{equation}
between the symmetric and the antisymmetric ground state energy.
For the latter, we specifically obtain the probability distribution
\begin{equation}
  P(\Delta E_0) = \frac{2}{\pi} \frac{\overline{\Delta E_0}}
  {(\Delta E_0)^2 + (\overline{\Delta E_0})^2} \label{cauchy}
\end{equation}
with
\begin{equation}
  \overline{\Delta E_0} = \frac{2\pi V_{\rm eff}^2}{N \Delta_c} \label{split}
\end{equation}
where $\Delta_c$ denotes the mean level spacing in the chaos at energy $E_0$.
This distribution is, strictly speaking, valid only for $\Delta E_0 \ll V_{\rm eff}$
and exhibits a cutoff at $\Delta E_0 \sim 2 V_{\rm eff}$, which ensures that the
statistical expectation value $\langle\Delta E_0\rangle = \int_0^\infty x P(x) dx$ does not diverge.

Since tunneling rates and their parametric variations are typically studied on
a logarithmic scale (i.e., $\log (\Delta E_0)$ rather than $\Delta E_0$ is plotted vs.\
$1/ \hbar$, see Figs \ref{fg:t1}--\ref{fg:t3} below ), the relevant quantity to be
calculated from Eq.~(\ref{cauchy}) and compared to quantum data is not the
mean value $\langle\Delta E_0\rangle$, but rather the average of the {\em logarithm} of 
$\Delta E_0$.
We therefore define our ``average'' level splitting $\langle\Delta E_0\rangle_g$ as the
{\em geometric} mean of $\Delta E_0$, i.e.
\begin{equation}
  \langle\Delta E_0\rangle_g \equiv \exp \left[ \left\langle \ln(\Delta E_0) \right\rangle \right]
\end{equation}
and obtain as result the scale defined in Eq.~(\ref{split}),
\begin{equation}
  \langle\Delta E_0\rangle_g = \overline{\Delta E_0} \, .
\end{equation}

This expression further simplifies for our specific case of periodically
driven systems, where the time evolution operator $U$ is modeled by the
dynamics under the effective Hamiltonian (\ref{eq:model}) over one period $\tau$.
In this case, the chaotic eigenphases $\varphi_j^\pm \equiv \mathcal{E}_j^\pm \tau / \hbar$ are
uniformly distributed in the interval $0 < \varphi_j^\pm < 2\pi$.
We therefore obtain
\begin{equation}
  \Delta_c = \frac{2 \pi \hbar}{N \tau}
\end{equation}
for the mean level spacing near $E_0$.
This yields
\begin{equation}
  \langle\Delta \varphi_0\rangle_g \equiv \frac{\tau}{\hbar} \langle\Delta E_0\rangle_g = \left( \frac{\tau V_{\rm eff}}{\hbar} \right)^2
  \label{eq:splitg}
\end{equation}
for the geometric mean of the ground state's eigenphase splitting.
Note that this final result does not depend on how many of the chaotic states
do actually participate in the sub-block $(H_{ij}^\pm)$;
as long as this number is sufficiently large to justify the validity of the
Cauchy distribution (\ref{cauchy}) (see Ref.~\cite{LeyUll96JPA}), the
geometric mean of the eigenphase splitting is essentially given by the
square of the coupling $V_{\rm eff}$ from the ground state to the chaos.

The distribution (\ref{cauchy}) also permits the calculation of the
logarithmic variance of the eigenphase splitting: we obtain
\begin{equation}
  \left\langle \left[ \ln(\Delta \varphi_0) - \langle \ln(\Delta \varphi_0) \rangle \right]^2 \right\rangle 
  = \frac{\pi^2}{4} \, . \label{eq:var}
\end{equation}
This universal result predicts that the actual splittings may be enhanced or
reduced compared to $\langle\Delta \varphi_0\rangle_g$ by factors of the order of $\exp(\pi/2) \simeq 4.8$,
independently of the values of $\hbar$ and external parameters.
Indeed, we shall show in the following section that short-range fluctuations
of the splittings, arising at small variations of $\hbar$, are well characterized
by the standard deviation that is associated with Eq.~(\ref{eq:var}).

\section{Application to the kicked Harper model}

\label{s:kh}

\begin{figure}
  \includegraphics*[width=\textwidth]{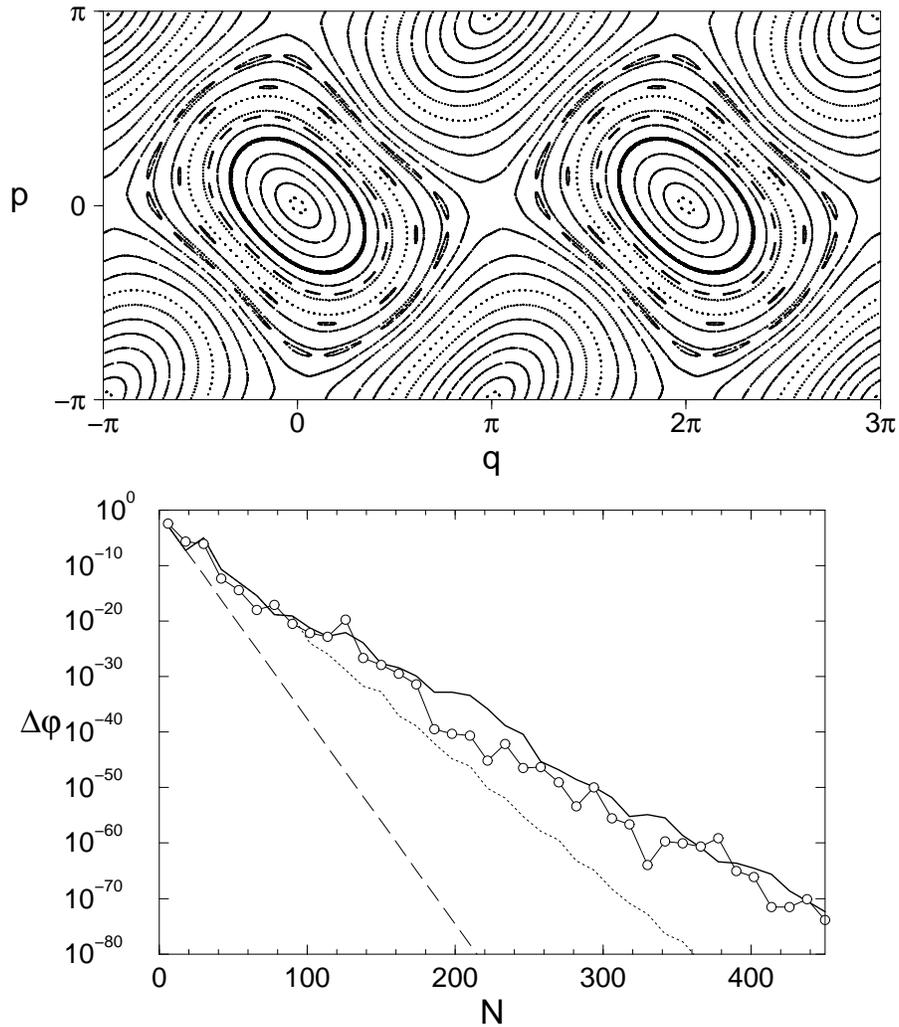}
  \caption{
    Classical phase space (upper panel) and quantum eigenphase splittings
    (lower panel) of the kicked Harper at $\tau=1$.
    The latter are calculated for the $n$th excited states at 
    $N \equiv 2\pi/\hbar = 6(2n+1)$ --- i.e., for all states that are semiclassically
    localized on the torus with action variable $\pi/6$.
    The decay of the exact quantum splittings (circles) is quite well
    reproduced by the semiclassical prediction (thick solid line) that is
    based on the 16:2, the 10:1, and the 14:1 resonance (highlighted in the
    upper panel, together with the above torus).
    The dotted and dashed lines show, respectively, the semiclassical
    splittings that are obtained by taking into account only the 10:1
    resonance, and the ``unperturbed'' splittings $\Delta\varphi_n^{(0)}$ that would 
    result from the integrable approximation (\ref{eq:h0}).
    \label{fg:t1}
  }
\end{figure}

To demonstrate the validity of our approach, we apply it to the 
``kicked Harper'' model \cite{LebO90PRL}
\begin{equation}
  H(p,q,t) = \cos p + \sum_{n=-\infty}^\infty \tau \delta( t - n \tau ) \cos q \, .
  \label{eq:kh}
\end{equation}
This model Hamiltonian is characterized by the parameter $\tau > 0$ which
corresponds to the period of the driving as well as to the strength of the
perturbation from integrability.
The classical dynamics of this system is described by the map $(p,q) \mapsto (p',q')
\equiv \mathcal{T}(p,q)$ with
\begin{eqnarray}
  p' & = & p + \tau \sin q \label{eq:map1} \\
  q' & = & q - \tau \sin p' \label{eq:map2}
\end{eqnarray}
that generates the stroboscopic Poincar{\'e} section at times immediately
before the kick.
The phase space of the kicked Harper is $2\pi$ periodic in position and
momentum, and exhibits, for not too large perturbation strengths $\tau$, a region
of bounded regular motion centered around $(p,q) = (0,0)$ (see the upper panel
of Fig.~\ref{fg:t1}).

The associated time evolution operator of the quantum kicked Harper is
given by
\begin{equation}
  U = \exp \left( - \frac{i \tau}{\hbar} \cos \hat{p} \right) 
  \exp \left( - \frac{i \tau}{\hbar} \cos \hat{q} \right)
\end{equation}
where $\hat{p}$ and $\hat{q}$ denote the position and momentum operator,
respectively.
The quantum eigenvalue problem drastically simplifies for $\hbar = 2 \pi / N$ with
integer $N > 0$, since the two phase-space translation operators 
$T_1 = \exp( 2 \pi i \hat{p} / \hbar )$ and $T_2 = \exp( -2 \pi i \hat{q} / \hbar )$
mutually commute with $U$ and with each other in that case \cite{LebO90PRL}.
This allows us to make a simultaneous Bloch ansatz in both position and
momentum --- i.e., to choose eigenstates with the properties 
\begin{eqnarray}
  \psi(q + 2\pi) & = & \psi(q) \exp(i\xi_q) \label{eq:bloch1} \\
  \hat{\psi}(p + 2\pi) & = & \hat{\psi}(p) \exp(i\xi_p) \label{eq:bloch2}
\end{eqnarray}
where $\hat{\psi}$ denotes the Fourier transform of $\psi$.
Since the subspace of wave functions satisfying
(\ref{eq:bloch1},\ref{eq:bloch2}) at fixed Bloch phases $\xi_q$ and $\xi_p$ has
finite dimension $N$, finite matrices need to be diagonalized to obtain the
eigenstates of $U$.

Quantum tunneling can take place between the central regular region around
$(0,0)$ and its periodically shifted counterparts.
The spectral manifestation of this classically forbidden coupling process is 
a finite bandwidth of the eigenphases $\varphi_n \equiv \varphi_n^{(\xi_q,\xi_p)}$ of $U$ that are
associated with the $n$th excited quantized torus within this region.
We shall not discuss this bandwidth in the following (the calculation of which
would require diagonalizations for many different values of $\xi_q$ and $\xi_p$),
but consider a simpler, related quantity, namely the difference
\begin{equation}
  \Delta \varphi_n = |\varphi_n^{(0,0)} - \varphi_n^{(\pi,0)}| \label{eq:khsplit}
\end{equation}
between the eigenphases of the periodic ($\xi_q = 0$) and the anti-periodic 
($\xi_q = \pi$) state in position, at fixed Bloch phase $\xi_p = 0$ in momentum.
In this way, we effectively map the tunneling problem to a double well
configuration, with the two symmetric wells given e.g.\ by the regions around
$(0,0)$ and $(2\pi,0)$.

Fig.~\ref{fg:t1} shows the eigenphase splittings of the kicked Harper in the
near-integrable regime at $\tau=1$.
The splittings were calculated for the $n$th excited states at $N = 6(2n+1)$,
i.e.\ for all possible states that are, in phase space, localized on the same
classical torus with action variable $\pi/6$.
These states were identified by comparing the overlap matrix elements of the
eigenstates of $U$ with the $n$th excited eigenstate (as counted from the
center of the region) of the time-independent Hamiltonian
\begin{eqnarray}
  H_0(p,q) & = & \cos p + \cos q - \frac{\tau}{2} \sin p \sin q \nonumber \\
  & & - \frac{\tau^2}{12} \left( \cos p \sin^2 q + \cos q \sin^2 p \right)
  - \frac{\tau^3}{48} \sin(2p) \sin(2q) \, , \label{eq:h0}
\end{eqnarray}
which represents a very good integrable approximation of $H$ at small $\tau$
\cite{BroSchUll02AP}.
Multiple precision arithmetics, based on the GMP library \cite{gmp}, was used
to compute eigenphase splittings below the ordinary machine precision limit.

The semiclassical calculation of the eigenphase splittings is based on three
prominent nonlinear resonances that are located between the torus with action
variable $\pi/6$ and the separatrix:
the 16:2 resonance \cite{remark}, the 10:1 resonance, and the 14:1 resonance.
Below $N \simeq 100$, we find that the tunneling process is entirely induced by
the 10:1 resonance (i.e., the most dominant one, according to the criterion
that $r$ and $s$ be minimal).
In this regime, the eigenphase splittings are reproduced by an expression of
the form (\ref{split1}) (with the resonance parameters $I_{r:s}$,
$m_{r:s}$, $V_{r:s}$ extracted from the classical phase space), where the
unperturbed splittings $\Delta \varphi_n^{(0)}$ are derived, via Eq.~(\ref{split0}),
from the imaginary action integrals $\sigma_n$ along complex orbits between the two
symmetry-related regular regions.

The other two resonances come into play above $N \simeq 100$, where they ``assist''
at the transition across the 10:1 resonance.
In that regime, a recursive application of the resonance-assisted coupling 
scheme, taking into account all possible perturbative pathways that involve
those resonances (see Ref.~\cite{BroSchUll02AP}), is applied to calculate the
semiclassical tunnel splittings.
We see that the result systematically overestimates the exact quantum
splittings above $N \simeq 200$, and does not properly describe their fluctuations.
We believe that this mismatch might be due to incorrect energy denominators
and coupling matrix elements that result from the simplified form (\ref{pend})
of the effective pendulum Hamiltonian.
The average exponential decay of the quantum splittings, however, is well
reproduced by the semiclassical theory.
A comparison with the unperturbed splittings $\Delta\varphi_n^{(0)}$ calculated from the
integrable approximation (\ref{eq:h0}) (dashed line in Fig.~\ref{fg:t1})
clearly demonstrates the validity of the resonance-assisted tunneling
mechanism.

\begin{figure}
  \includegraphics*[width=\textwidth]{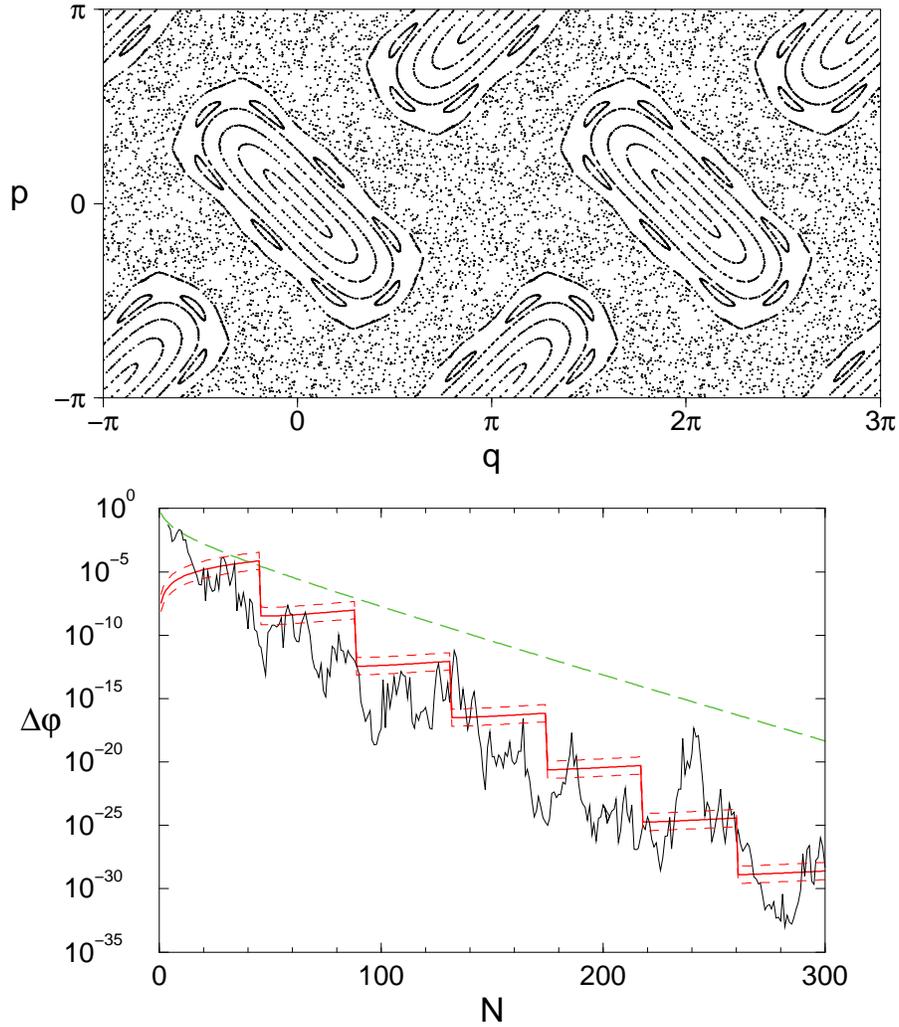}
  \caption{
    Classical phase space (upper panel) and quantum eigenphase splittings
    (lower panel) of the kicked Harper at $\tau=2$.
    The splittings are calculated for the ground state within the regular
    island, as a function of $N \equiv 2\pi/\hbar$.
    The thick solid line in the lower panel represents the semiclassical
    prediction of the eigenphase splittings, which is based on the 8:2
    resonance displayed in the upper panel.
    The size of the logarithmic standard deviation according to
    Eq.~(\ref{eq:var}) is indicated by the two dashed lines that accompany the
    semiclassical splittings; these dashed lines are defined by 
    $\langle\Delta \varphi_0\rangle_g \times \exp(\pm \pi/2)$.
    The long-dashed curve represents the prediction of the eigenphase
    splittings according to the theory of Ref.~\cite{PodNar03PRL} 
    (see Eq.~(\ref{eq:pod2})).
    \label{fg:t2}
  }
\end{figure}

At $\tau=2$, the phase space of the kicked Harper becomes mixed regular-chaotic,
and the regular region around $(0,0)$ turns into an island that is embedded
into the chaotic sea.
Fig.~\ref{fg:t2} shows the phase space together with the corresponding
eigenphase splittings.
The latter were calculated here for the ground state of the island, with the
dimension $N$ ranging from 4 to 300 in integer steps.
We clearly see that the fluctuations of the splittings are much more
pronounced than in the near-integrable regime.

The semiclassical calculation of the eigenphase splittings is based on a
prominent 8:2 resonance within the island (visible in the upper panel of
Fig.~\ref{fg:t2}) and evaluated according to Eq.~(\ref{eq:splitg}).
The sharp steps of $\langle\Delta \varphi_0\rangle_g$ arise from the artificial separation between
perfect regularity inside and perfect chaos outside the island:
when $\hbar$ drops below the value at which the $(kr)$th excited state of the
island is exactly localized on the outermost invariant torus, $(k+1)$ instead
of $k$ perturbative steps are required to connect the ground state to the
chaotic domain, and the corresponding coupling matrix element $V_{\rm eff}$
(\ref{eq:veff}) acquires an additional factor.
In reality, the transition to the chaos is ``blurred'' by the presence of
high-order nonlinear resonances and ``Cantori''
\cite{MacMeiPer84PRL,MeiOtt85PRL} in the vicinity of the island.
The latter provide efficient barriers to the quantum flow 
\cite{GeiRadRub86PRL,MaiHel00PRE,KetO00PRL} and therefore change the
structure of the block $(H_{ij}^\pm)$ in the matrix (\ref{eq:model}).
Hence, except for the case of a very ``clean'', structureless chaotic sea,
the steps are not expected to appear in this sharp form in the actual, quantum
splittings.

It is therefore remarkable that the quantum splittings exhibit plateaus at
approximately the same levels that are predicted by the resonance-assisted
mechanism.
These plateaus, however, seem to be ``shifted'' to the left-hand side with
respect to the semiclassical splittings --- i.e., the latter apparently
overestimate the position of the steps.
We believe that this is due to a rich substructure of partial barriers within
the chaos, which effectively increases the phase space area of the region in
which quantum transport is inhibited.
Such a substructure is not visible in the Poincar\'e surface of section, but its
existence becomes indeed apparent when individual trajectories are propagated
in the vicinity of the island.
A more refined approach, taking into account such partial barriers in the
chaos, would probably be required to obtain a better agreement.

\begin{figure}
  \includegraphics*[width=\textwidth]{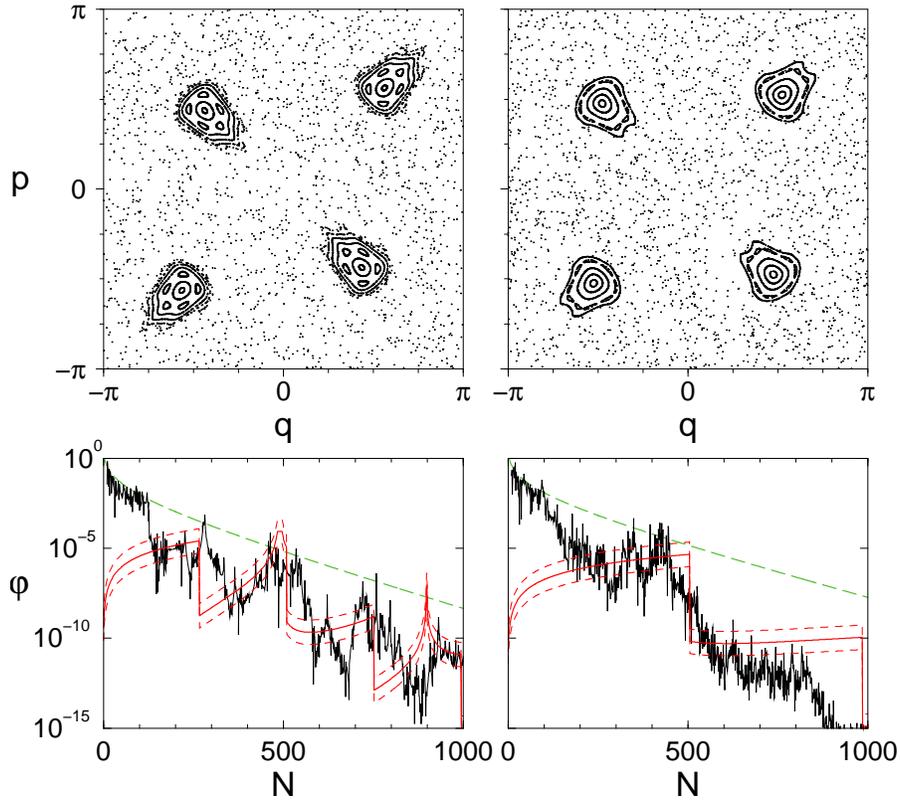}
  \caption{
    Chaos-assisted tunneling in the kicked Harper at $\tau = 2.8$ (left column)
    and $\tau = 3$ (right column).
    Calculated are the eigenphase splittings between the symmetric and the
    antisymmetric quasimodes that are localized on the two bifurcated islands
    (i.e., the two islands facing each other on the $q = -p$ axis),
    for periodic boundary conditions $\xi_q = \xi_p = 0$.
    The semiclassical calculation (thick lines in the lower panels) is based
    on a 5:4 resonance at $\tau = 2.8$ and on a 9:8 resonance at $\tau = 3$.
    The huge peaks in the lower left panel are induced by internal
    near-resonant transitions within the island, which are permitted since the
    5:4 resonance lies rather closely to the center of the island.
    As in Fig.~\ref{fg:t2}, the two dashed lines accompanying the
    semiclassical splittings indicate the size of the logarithmic standard
    deviation according to Eq.~(\ref{eq:var}), and the long-dashed curve
    represents prediction according to the theory of Ref.~\cite{PodNar03PRL}.
    \label{fg:t3}
  }
\end{figure}

The central fixed point $(0,0)$ becomes unstable at $\tau=2$, when the 2:1
resonance emerges in the center of the island.
This 2:1 resonance develops into a symmetric pair of regular islands located
along the $p = -q$ axis, which dominate the phase space for $\tau > 2.5$
together with their counterparts from the bifurcation of the fixed point
at $(\pi,\pi)$.
The stable periodic points of the 2:1 resonance are fixed points of the
twicely executed kicked Harper map $\mathcal{T}^2$ 
(i.e., where Eq.~(\ref{eq:map1},\ref{eq:map2}) is applied twice).
Therefore, quantum states associated with those islands appear, at fixed Bloch
phases $\xi_q, \xi_p$, as doublets in the eigenphase spectrum of the corresponding
time evolution operator $U^2$.
We shall show now that the splitting between the levels of such a doublet is
again described by the resonance- and chaos-assisted tunneling scenario in the
semiclassical regime.

The doublets associated with the ground state of the 2:1 resonance islands are
calculated by diagonalizing $U^2$ and by identifying those eigenstates that
most sharply localized around the centers of the two islands
(we restrict ourselves to the pair that is located on the $q=-p$ axis).
Fig.~\ref{fg:t3} shows the corresponding eigenphase splittings as a function
of $N = \hbar / 2\pi$, computed for periodic boundary conditions $\xi_q = \xi_p = 0$
at $\tau = 2.8$ (left column) and $\tau = 3$ (right column).
The semiclassical calculation of the eigenphase splittings is performed in the
same way as for $\tau=2$, with the little difference that $2\tau$ rather than $\tau$ is
used as period in Eq.~(\ref{eq:splitg}).
For the coupling to the chaos, we identify a relatively large 5:4
sub-resonance within the 2:1 resonance islands at $\tau = 2.8$, and a smaller
9:8 sub-resonance at $\tau=3$.

It is instructive to notice that a rather small variation $\Delta \tau = 0.2$ of the
perturbation parameter can lead to qualitatively different features of the
tunneling rates (compare the lower panels of Fig.~\ref{fg:t3}), and that these
features are indeed reproduced by the resonance-assisted tunneling scheme.
At $\tau = 2.8$, the 5:4 sub-resonance is located sufficiently closely to the
center of the island that it induces {\em near-resonant internal transitions}
within the island , i.e.\ couplings from the ground state to the $(5l)$th
excited states that are strongly enhanced in the expression (\ref{eq:veff})
due to nearly vanishing energy denominators $E_0 - E_{5l}$.
As a consequence, pronounced peaks in the tunneling rate are obtained in the
vicinity of such internal near-degeneracies \cite{remark2}.
Such peaks are indeed displayed by the exact quantum splittings as well,
though not always at exactly the same position and with the same height as
predicted by semiclassics.
A completely different scenario is encountered at $\tau=3$, where the coupling to
the chaos is mediated by a tiny 9:8 sub-resonance that is closely located to
the outermost invariant torus of the island.
This high-order resonance induces rather larges plateaus in the chaos-assisted
tunneling rates, which are clearly manifest in the quantum splittings, and
which indicate that the coupling to the chaos is essentially governed by the
same matrix element over a wide range of $\hbar$.

In addition to the resonance-assisted eigenphase splittings (\ref{eq:splitg}),
we also plot in Figs.~\ref{fg:t2} and \ref{fg:t3} the prediction that is based
on the semiclassical expression (\ref{eq:pod}) proposed by Podolskiy and
Narimanov \cite{PodNar03PRL}.
This expression involves the unknown rate $\gamma$ which, however, is independent
of $\hbar$ and characterizes the tunneling process also in the deep anticlassical
regime $\hbar \sim 1$.
Hence, we conclude that, for dimensionality reasons, $\gamma$ has to be of the
order of the intrinsic scale $\tau^{-1}$ of our system, up to dimensionless
prefactors of the order of unity.
We therefore set $\gamma \tau \equiv 1$ and obtain
\begin{equation}
  \Delta \varphi \simeq \frac{\Gamma(\nu,2\nu)}{\Gamma(\nu+1,0)} \simeq \frac{1}{\sqrt{2\pi \nu^3}} {\rm e}^{-(1-\ln 2) \nu} 
  \label{eq:pod2}
\end{equation}
as prediction for the average eigenphase splittings, with $\nu = A / (\pi \hbar)$
where $A$ is the phase space area covered by the island.
In all cases that were studied in this work, we find good agreement of
Eq.~(\ref{eq:pod2}) with the exact quantum splittings for comparatively low 
$1 / \hbar$, and significant deviations deeper in the semiclassical regime.
This indicates that the simple harmonic-oscillator approximation for the
dynamics within the island (which is needed in order to predict the tunneling
tail of the eigenfunction) is correct for large $\hbar$, but becomes invalid as
soon as nonlinear resonances come into play.

While the exact quantum splittings may, depending on the size of $\hbar$,
considerably deviate from both semiclassical predictions (\ref{eq:splitg}) and
(\ref{eq:pod2}) of the average, the short-range fluctuations of the splittings
seem to be well characterized by the universal expression (\ref{eq:var}) for
the logarithmic variance:
As is shown by the dashed lines accompanying $\langle\Delta \varphi_0\rangle_g$ in Figs.~\ref{fg:t2},
the amplitudes of those fluctuations are more or less contained within the
range that is defined by the standard deviation associated with
Eq.~(\ref{eq:var}).
This confirms the general validity of the chaos-assisted tunneling scenario.


\section{Conclusion}

\label{s:c}

In summary, we have presented a straightforward semiclassical scheme to
reproduce tunneling rates between symmetry-related regular islands in mixed
systems.
Our approach is based on the presence of a prominent nonlinear resonance
which induces a coupling mechanism between regular states within and chaotic
states outside the islands.
The associated coupling matrix element can be directly extracted from
classical quantities that are associated with the resonance.
Assuming the presence of a structureless chaotic sea, a random matrix ansatz
can be made for the chaotic part of the Hamiltonian, which results in a simple
expression for the average tunneling rate in terms of the above matrix
element.

Application to the kicked Harper model shows good overall agreement and
confirms that plateau structures in the quantum splittings originate indeed
from the influence of nonlinear resonances.
A significant overestimation of the quantum splittings is generally found for
``weakly'' chaotic systems where a large part of the phase space is covered by
regular islands.
This is tentatively attributed to the presence of a rich substructure of
partial barriers (such as Cantori or island chains) in the chaotic sea.
Such partial barriers generically manifest in the immediate vicinity of
regular islands and inhibit, for not too small $\hbar$, the quantum transport in a
similar way as invariant tori \cite{GeiRadRub86PRL,MaiHel00PRE}.
Semiclassical studies in the annular billiard \cite{DorFri95PRL,FriDor98PRE}
do indeed indicate that quantum states localized in this hierarchical region
\cite{KetO00PRL} play an important role in the dynamical tunneling process
\cite{rem_ab}.

In the ``anticlassical'' regime of large $\hbar$, our semiclassical theory
systematically underestimates the exact quantum splittings.
This might be due to the approximations that are involved in the present
implementation of the resonance-assisted tunneling scheme (we neglect, e.g.,
the action dependence of the Fourier coefficients of the effective potential
in Eq.~(\ref{eq:Vres}), which could play an important role in this regime).
It is also possible that the coupling to the chaos is induced by a 
{\em different} mechanism at large $\hbar$, which effectively amounts to
extracting the associated matrix element from the overlap of the tunneling
tail of the local semiclassical wave function with the chaotic phase space.
Indeed, we find that the simple semiclassical expression (\ref{eq:pod})
introduced by Podolskiy and Narimanov \cite{PodNar03PRL}, which is essentially
based on that scheme, reproduces the tunneling rates quite well in this
regime.

The validity of the resonance-assisted tunneling mechanism was confirmed not
only for the kicked Harper, but also for the kicked rotor \cite{EltSch05PRL}
as well as for the driven pendulum Hamiltonian that describes dynamical
tunneling of cold atoms in periodically modulated optical lattices
\cite{MouO01PRE}.
The theory can be furthermore generalized to describe chaos-assisted decay
processes of quasi-bound states in open systems, such as the ionization of
non dispersive wave packets in microwave-driven hydrogen \cite{WimO05}.
Ongoing studies on dynamical tunneling in autonomous model Hamiltonians with
two and three degrees of freedom \cite{Kes05JCP,Kes05XXX} clearly reveal that
nonlinear resonances play an equally important role in more complicated
systems as well.
The mechanism presented here might therefore provide a feasible scheme to
predict, understand, and possibly also control dynamical tunneling in a
variety of physical systems.

\section*{Acknowledgement}

It is a pleasure to thank E.~Bogomolny, O.~Bohigas, O.~Brodier,
A.~Buchleitner, D.~Delande, S.~Fishman, S.~Keshavamurthy, P.~Leboeuf,
A.M.~Ozorio de Almeida, and S.~Tomsovic for fruitful and inspiring
discussions.


\begin{thebibliography}{10}
\bibitem{rem_compl}
This assumes, of course, that the double well potential $V$ is analytic.
\bibitem{rem_ad}
This step involves, strictly speaking, another time-dependent canonical
  transformation $(I,\vartheta) \mapsto (\widetilde{I},\widetilde{\vartheta})$
  which slightly modifies $I$ and $\vartheta$ (see also
  Ref.~\cite{BroSchUll02AP}).
\bibitem{remark}
Due to the internal symmetries of the kicked Harper, all $r$:$s$ resonances
  with relative prime $r,s$ and $r$ being a multiple of 4 have $2r$ instead of
  $r$ islands. To be consistent with Eq.~(\ref{pend}), we name them $2r$:$2s$
  resonances.
\bibitem{remark2}
In the numerical evaluation of the expression (\ref{eq:veff}), we took care
  that the admixture of the $(rl)$th excited state to the ground state may not
  exceed the upper limit 1, for all $ l = 1 \ldots (k-1)$. The prominent peak
  at $N \simeq 500$ in the lower panel of Fig.~\ref{fg:t3} is therefore
  rounded.
\end{thebibliography}
\end{document}